\documentclass[pra, twocolumn,showpacs]{revtex4}  % twocolumn preprint
\usepackage{graphicx}
\usepackage{amssymb}

\def\d#1{#1^\dagger}
 
\def\ket#1{|#1\rangle}

\begin{document}

\title{Graph states and carrier-envelope phase squeezing}
\author{Olivier Pfister}
\affiliation{Department of Physics, University of Virginia, 382 McCormick Road, Charlottesville, VA 22904-4714, USA}
\begin{abstract}
We outline a new class of continuous-variable graph states that can be useful to describe entanglement, and also multimode squeezing, in an optical frequency comb. We show that a particular case of such states coincides with the squeezing of the carrier-envelope phase, or of the total intensity, of a mode-locked laser. We then discuss the experimental issues related to measuring the quantum noise of the carrier-envelope phase of a frequency comb and show that this can be carried out by use of quantum heterodyne multiplexing, a technique that may be useful to measure multipartite entanglement as well.
\end{abstract}
\maketitle

\paragraph{Introduction.} 
The description of multipartite entanglement by graph states \cite{hein:2006} was recently developed in conjunction with the one-way quantum computing model, in which entangled states of the cluster type are used \cite{raussendorf:2001}. As the name indicates, a graph quantum state is completely described by a mathematical graph, i.e.\ a set of vertices connected by edges \cite{hein:2006} as per precise conventions. A vertex represents a physical system, e.g.\ a qubit (or qubit, 2-dimensional Hilbert space) qudit ($d$-dimensional Hilbert space), or qunat (continuous Hilbert space) \cite{lloyd:1999}, in a specified quantum state. An edge between two vertices represents the physical interaction between the corresponding systems. In this article, we will use different graph definitions, i.e. conventions for edges and vertices, from those used for cluster states. We will also use continuous quantum variables, that is, position and momentum, or quadrature amplitudes of the quantized electromagnetic field (which are equivalent to the former by virtue of the quantum harmonic oscillator formalism). Note that continuous-variable (CV) entanglement can always be recast in terms of discrete variables in an infinite dimensional Hilbert space by use of the Schmidt decomposition \cite{vanenk:1999,eberly:2003}, hence qunats can always be viewed as qudits.

Our group proposed a compact and scalable experimental implementation of Greenberger-Horne-Zeilinger (GHZ) states by use of multimode squeezing generated by {\em concurrent} optical nonlinearities in a single, sophisticatedly engineered, nonlinear optical medium \cite{pfister:2004,bradley:2005}. A demonstration of such a nonlinear medium was subsequently made in periodically poled potassium titanyl phosphate \cite{pooser:2005}. More recently, we showed theoretically that the generation of CV cluster states in a single optical parametric oscillator (OPO) is possible and gave a detailed experimental proposal for a square cluster \cite{menicucci:2007}.  Our method can be described by simple graph states, as explained below. In this paper, we extend the application of such graph states to multimode squeezing and then focus on the squeezing of the total phase of a set of coherent fields, regularly spaced in frequency (``frequency comb"), such as emitted by a mode-locked pulsed laser. This total phase is also called the carrier-envelope phase and its control is crucial in the use of femtosecond-laser frequency combs for defining frequency standards \cite{jones:2000,udem:1999}. We then describe experimental procedures for generating and for measuring, by quantum heterodyne multiplexing, the squeezing of the carrier-envelope phase. The latter being a CV GHZ entanglement witness, these procedures can be used for entanglement measurements as well.

\paragraph{Two-mode-squeezed graph states.}
The generation of scalable CV GHZ states was first proposed by van Loock and Braunstein \cite{vanloock:2000} and demonstrated experimentally for 3 modes \cite{jing:2003,aoki:2003}. In this method, $N$ single-mode squeezed states are generated by $N$ independent optical parametric amplifiers (OPA) and subsequently interfered at a $2N$-port beam splitter network with all optical paths controlled. A similar setup can be used to generate linear \cite{zhang:2006} or even arbitrary \cite{vanloock:2006} cluster states. Such proposals nonetheless lead to nontrivial, if polynomial, experimental complexity when scaling to large $N$. In contrast, our approach uses a {\em single} OPA, or OPO, as a multimode squeezer that delivers $N$ entangled modes directly, provided the nonlinear medium constituting the OPA phasematch simultaneously all two-mode interactions between every one of the optical modes \cite{pfister:2004}. This lends itself quite naturally to a graph description. 
%------------
\begin{figure}[htb]
\begin{center}
\begin{tabular}{c}
\includegraphics[width= .75 \columnwidth]{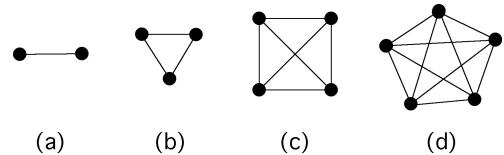}
\end{tabular}
\end{center}
%\vspace{-.3 cm}
\caption{GHZ graph states with two-mode squeezing interactions only. The $N$ modes are denoted by the vertices, connected by $N(N-1)/2$ interaction edges. (a) $N=2$, (b) $N=3$, (c) $N=4$, (d) $N=5$.}
\label{fig:graphs}
\end{figure} 
%------------
As an example, we consider GHZ states depicted in Fig.\ \ref{fig:graphs}. These are created by the second-order nonlinear interaction Hamiltonian \cite{pfister:2004}
\begin{equation}
H_1 = i\hbar\kappa\sum_{n<p} \d a_n\d a_p + H.c.
\end{equation}
where $\kappa$ is the nonlinear coupling constant and each term is represented by an edge of the graph. Each mode is a vertex and a qudit in the photon-number basis or a qunat in the quadrature amplitude basis \cite{vanenk:1999}, labeled by frequency as well as the wave vector and polarization directions. The set of interacting modes is defined by the resonant modes of the OPO optical cavity, inside which the nonlinear concurrent emitter is placed. It is straightforward to see from the Heisenberg equations for the modes $\{a_n\}_{1\dots N}$ that the matrix describing this system of coupled differential equations is proportional to the {\em adjacency matrix} \cite{hein:2006} of the graph so defined. In this case of a GHZ state, the graph is fully connected and the adjacency matrix ${\cal A}$ has elements ${\cal A}_{kl} = 1 - \delta_{kl}$, where $\delta$ is the Kronecker symbol, has eigenvalues $\{ N-1,-1,\dots,-1\}$ and respective (unnormalized) eigenvectors $\{(1,1,\dots, 1)$, $(1,-1,0,\dots, 0)$, $(1,0,-1,0,\dots, 0)$, \ldots, $(1,0,\dots, 0,-1)\}$ \cite{pfister:2004}. In terms of multimode squeezing of the field quadratures $X_j=a_j+\d a_j$ (``amplitude") and $P_j=i(\d a_j-a_j)$ (``phase"), this eigensystem yields the following solution to the quantum evolution: the phase sum $\mathbf{P}=\sum_{j=1}^N P_j$ is squeezed $\mathbf{P}(t)=\mathbf{P}(0) e^{-(N-1)\kappa t}$ (note the squeezing enhancement factor $N-1$), along with all amplitude differences $X_{kl}=X_k-X_l$, $X_{kl}(t)=e^{-\kappa t}X_{kl}(0)$ \cite{note}. In the limit of infinite squeezing, this means the result of the quantum evolution is a common eigenstate of all these operators, which coincide with the entanglement witnesses of a GHZ state \cite{vanloock:2003a,braunstein:2005a}. Such a state has the unnormalized expressions, in the amplitude-quadrature basis,
\begin{equation}
\int\ \ket x_1\ket x_2 \cdots \ket x_N\ dx.
\end{equation}
We have named this particular topology of nonlinear interactions a \emph{nonlinear optical concurrence (NOC)} --- not to be mistaken for the concurrence as a mathematical measure of entanglement \cite{hill:1997} --- because it mandates the co-occurrence of several nonlinear interactions (edges) for the same mode (vertex). The $N=2$ entangling NOC has a trivial graph and is already well-known from previous studies of the Einstein-Podolsky-Rosen (EPR) paradox: the corresponding entangled EPR state can, indeed, be generated from either a single nondegenerate OPA \cite{ou:1992} or from the interference of two single-mode OPAs \cite{furusawa:1998,lugiato:2002}. 

\begin{figure}[htb]
\begin{center}
\begin{tabular}{c}
\includegraphics[width= .75 \columnwidth]{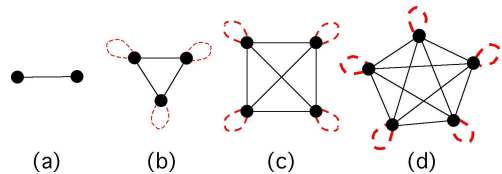}
\end{tabular}
\end{center}
%\vspace{-.3 cm}
\caption{GHZ graph states with single-mode (degenerate, loops) as well as two-mode (nondegenerate, 2-vertex edges) squeezing interactions. Here the loops all correspond to interactions opposite to those of the edges, i.e.\ downconverting versus upconverting, or vice versa.}
\label{fig:graphs2}
\end{figure}

Implementing an entangling NOC would seem a fairly involved task, for reasons we detailed previously \cite{pfister:2004,bradley:2005}, but we have proven it feasible \cite{pooser:2005}. One issue is that CV GHZ graphs cannot have any self-loops of the {\em same} type of interaction lest entanglement be lost. Indeed, as we previously established \cite{pfister:2004}, the protocol of van Loock and Braunstein \cite{vanloock:2000} corresponds to the Hamiltonian
\begin{equation}
H_2 = i\hbar\kappa\left(\sum_{n<p} \d a_n\d a_p - \frac{N-2}2 \sum_n \frac{a_n^{\dagger 2}}2\right)+ H.c.,
\end{equation}
which corresponds to a self-loop graph because of the second, degenerate interaction term. However, the graph is now weighted: all self-loops describe an interaction of opposite sign and different strength ($-(N-2)/2$) with respect to the 2-vertex edges (Fig.\ \ref{fig:graphs2}). 
%------------
%------------
Experimentally, it means that, if edges are parametric downconverting interactions, then the self-loops must be upconverting ones, or vice versa. Such a configuration is clearly incompatible with a single OPO, which is why the NOC scheme of Fig.\ \ref{fig:graphs} is interesting for the single-OPO implementation.

\begin{figure}[htb]
\begin{center}
\begin{tabular}{c}
\includegraphics[width= .75 \columnwidth]{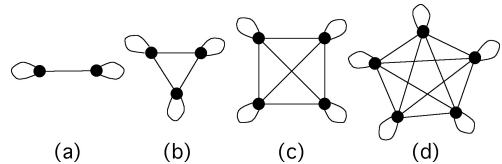}
\end{tabular}
\end{center}
%\vspace{-.3 cm}
\caption{Graph states for total-phase or total-amplitude squeezing.} %These states do not necessarily possess any entanglement.}
\label{fig:graphs3}
\end{figure}

If one now changes the signs of the self-loops, however, the Hamiltonian becomes 
\begin{equation}
H_3 = i\hbar\kappa\left(\sum_{n<p} \d a_n\d a_p + \sum_n \frac{a_n^{\dagger 2}}2\right)+ H.c.,
\end{equation}
with the corresponding graphs given in Fig.\ \ref{fig:graphs3}.
%------------
%------------
The adjacency matrix has all its elements equal to 1 and has eigenvalues $\{N,0,\cdots,0\}$ with the same respective eigenvectors. The only squeezed operator here is the phase sum $\mathbf{P}(t)=\mathbf{P}(0) e^{-N\kappa t}$ (note the squeezing enhancement factor $N$). The other entanglement witnesses, the amplitude differences, are now constants of the motion (zero eigenvalue), that is, not squeezed and hence no longer diagonalized by the resulting state in the limit of infinite squeezing. %Moreover, $H_3$ is not sufficient to obtain any entanglement: an example of phase-sum eigenstate is $\ket{p_1}_1\ket{p_2}_2\cdots\ket{p_n}_n$.

\begin{figure}[htb]
\begin{center}
\begin{tabular}{c}
\includegraphics[width= .9 \columnwidth]{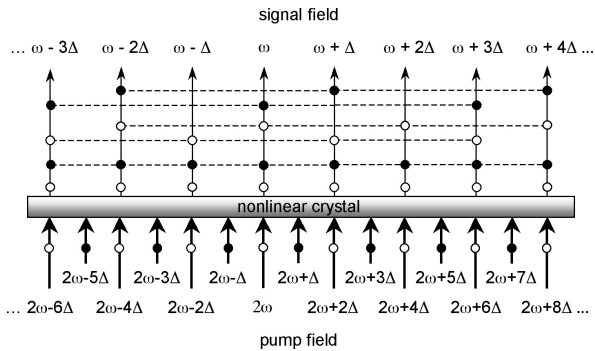}
\end{tabular}
\end{center}
%\vspace{-.3 cm}
\caption{Experimental implementation of the graph states such as Fig.\ \ref{fig:graphs3}. All pump field spectral components have the same polarization. The open and filled circles label, respectively, the pump components who have a degenerate interaction with the signals, or not. For clarity, not all signal couplings have been drawn.}
\label{exp}
\end{figure}

\paragraph{A phase-sum squeezer: the broadband degenerate OPO.}
The Hamiltonian $H_3$ is actually very easy to implement, compared to $H_{1,2}$: it describes an OPA or OPO pumped by a train of mode-locked pump pulses whose repetition rate is equal to the free spectral range of the OPA cavity, as sketched in Fig.\ \ref{exp}.
%------------
%------------
It is then easy to see that each pump frequency in Fig.\ \ref{exp} contributes one skew diagonal of ones to the adjacency matrix, thereby yielding the aforementioned uniform adjacency matrix, corresponding to the Heisenberg system for $H_3$. Note that the match between the pump's repetition rate and the OPO cavity free spectral range is crucially important but can be easily controlled by use of an electronic servo loop. For pump and signal spectral ranges within the phase-matching bandwidth of the OPO crystal, one should therefore obtain phase-sum squeezing of the signal comb. (Note that the OPO cavity linewidth is typically much narrower than the phase-matching bandwidth and than the free spectral range.)

In the case of a broad femtosecond frequency comb whose high-frequency end of the spectrum would be the pump modes, and whose low-frequency end of the spectrum would be the signal modes, it is then not clear whether one would obtain carrier-envelope phase squeezing since only the sum of the signal phases would be squeezed, according to the analysis of this paper. However, one must keep in mind that this paper assumes undepleted classical pump modes, which is not necessarily true in that situation. One should then include the quantum behavior of the pump fields during the interaction, using more sophisticated quantum optical methods \cite{wm}. Such a study is compelling but out of the scope of the present paper. Note that a complementary mathematical analysis of the same system was carried out independently of this work \cite{Valcarcel} but the squeezed physical observables were called ``supermodes" and not identified.

\begin{figure}[hbtf]
\begin{center}
\begin{tabular}{c}
\includegraphics[width=0.75\columnwidth]{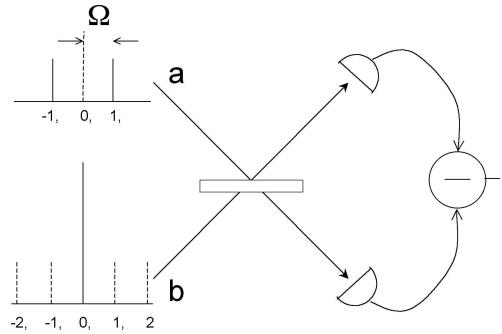}
\end{tabular}
\end{center}
\vspace{-.3 cm}
\caption{Detection of total-phase squeezing. The $a$ input port receives the 2 modes whose phase-sum is squeezed. The $b$ mode receives a coherent LO field. The vacuum modes that contribute to the signal at frequency $\Omega$ are identified by dashed lines. The 2 vacuum modes at the extremities of the $b$ input spectrum are the ``image" modes.}
\label{het}
\end{figure}

\paragraph{Phase-sum squeezing detection: quantum heterodyne multiplexing.} 
Considering again the case of independent pump and signal combs, we now show that the detection of the signal phase-sum can be achieved relatively easily by balanced heterodyne detection.  Heterodyne detection does not usually allow squeezing to be measured because it lets in vacuum fluctuations. However, there exists a particular case where this effect can be occulted. This case is the balanced heterodyne detection of two frequency nondegenerate modes by use of a single local oscillator (LO) field of frequency the exact average of the modes' frequencies. 
This is the method that was used in the first squeezing measurement \cite{slusher:1985}. In the photocurrent difference signal, all DC signals are filtered out, as well as the interference signals between modes that impinge onto the beam splitter through the same input port. We consider the configuration where all quantum squeezed modes are input into the same beamsplitter port and the local oscillator is sent into the other port, as depicted in Fig.\ \ref{het}.
%------------
%------------

If we denote by $a_j$ the annihilation operator of the $a$ input mode at frequency $\omega +j\Omega$, and $b_j$ that of the $b$ input mode at frequency $\omega +j\Omega$, the operator detected by the setup of Fig.\ \ref{het} after photocurrent subtraction is 
\begin{eqnarray}
S = \d ab+a\d b & \equiv & \sum_j \d a_j \sum_k b_k + \sum_j  a_j \sum_k \d b_k \\
& \simeq & (\d a_{-1}+\d a_1)\ b_0 +  (a_{-1}+ a_1)\ \d b_0 \\
& \simeq & \beta_0\ A^{(1+)}_{\theta}
\end{eqnarray}
in the limit of a classical LO whose amplitude is much larger than those of the squeezed fields, in which case the beats of the $a$ fields with the image bands in $b$ can be neglected. We have posed $b_0=\beta_0e^{i\theta}$ a complex number, the quadrature $A^{(j)}_{\theta}=e^{-i\theta}a_j+e^{i\theta}\d a_j$, and $a^{(j+)}=a_j+a_{-j}$. For the LO phase $\theta=\pi/2$, one will measure, at photocurrent frequency $\Omega$, the sum of the signal phases $A^{(1+)}_{\pi/2}\equiv P_1+P_{-1}$.
%------------
\begin{figure}[htb]
\begin{center}
\begin{tabular}{c}
\includegraphics[width=0.75\columnwidth]{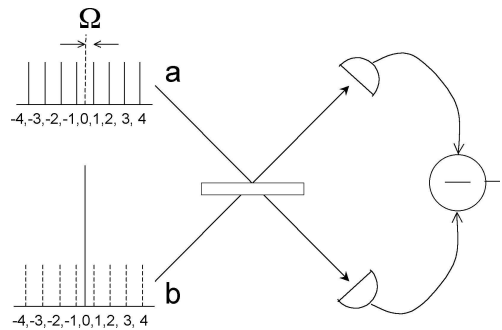}
\end{tabular}
\end{center}
\vspace{-.3 cm}
\caption{Detection of total-phase squeezing for a frequency comb.}
\label{hetm}
\end{figure} 
%------------

It is straightforward to extend the original two-mode measurement to a larger number of modes by making use of the frequency degree of freedom of the photocurrent: in Fig.\ \ref{hetm}, each mode pair $a_{\pm j}$ of the signal comb is beat against the LO, yielding a signal at $(2j-1)\Omega$ in the photocurrent difference. By virtue of the properties of balanced heterodyne detection stated above, demodulating the photocurrent difference at all such frequencies and summing all results will yield the operator
\begin{equation}
S = \beta_0 \sum_{j=1}^n A^{(j+)}_{\theta} =  \beta_0 \sum_{j=-n}^n A^{(j)}_{\theta},
\end{equation}
which coincides with the total phase $\mathbf P$ for $\theta=\pi/2$. In the expression above, note that $n$ ideally equals the total mode number ($N$) in the squeezed operator. In practice, $n\ll N$ is possible due to a detection bandwidth limit of $n\Omega$. This would create a problem since the fact that $\sum_{j=-N}^N P_j$ is squeezed does not imply that $\sum_{j=-n}^n P_j$ is squeezed. This problem can easily be surmounted, however, by the use of a comb of several LO fields, spaced in frequency by $n\Omega$ (assuming the detection band-shape is a rectangular window). The total output signal can then be demodulated and summed in exactly the same manner to yield the same result. Overall, this use of the frequency domain for encoding quantum information more densely is a quantum analog of FM radio. We name this detection method ``quantum heterodyne multiplexing." 

In conclusion, we have shown that the topology of graph states, defined with squeezing rather than controlled-phase interactions \cite{hein:2006}, connects to entanglement and multimode squeezing, and in particular phase-sum squeezing, of interest to improve the metrological performance of frequency combs. Further theoretical study is needed at this point to investigate the possibilities for the self-referenced metrological comb. Finally, we proposed a compact multimode detection method for the total phase, which can be useful for measurements of multipartite entanglement as well. This work was supported by NSF Grants No.\ PHY-0555522 and No.\ CCF-0622100.

\end{document}